\documentstyle{article}

\font\tenrm=cmr10
\font\tenit=cmti10
\font\elevenbf=cmbx10 scaled\magstep 1
\font\elevenrm=cmr10 scaled\magstep 1
\font\elevenit=cmti10 scaled\magstep 1

\renewenvironment{thebibliography}[1]
 { \elevenrm
   \begin{list}{\arabic{enumi}.}
    {\usecounter{enumi} \setlength{\parsep}{0pt}
     \setlength{\itemsep}{3pt} \settowidth{\labelwidth}{#1.}
     \sloppy
    }}{\end{list}}

\begin{document}

\baselineskip=13pt

\rightline {\bf UG-FT-55/96}
\rightline {hep-ph/9603347}
\rightline {\elevenrm March \ 1996}

\begin{center}
\vglue 1.5cm
{
 {\elevenbf        \vglue 10pt
     Search for new neutral bosons at future colliders
               \footnote{
\tenrm
\baselineskip=11pt
Lecture delivered at the International Conference 
{\tenit "Particle Physics in the Standard Model and Beyond"},
Bystra, September 19-26, 1995\\}
\vglue 5pt}

\vglue 1.5cm
{\tenrm F. DEL AGUILA, M. MASIP, M. P\'EREZ-VICTORIA\\}
\baselineskip=13pt
\vglue 0.2cm
{\elevenit Departamento de F\'\i sica Te\'orica y del Cosmos, 
Universidad de Granada \\}
\baselineskip=12pt
{\elevenit Granada, 18071, Spain \\}}
\vglue .5cm

\centerline{\elevenit (to appear in Acta Physica Polonica)}

\vglue 1.5cm
{\tenrm ABSTRACT}

\end{center}

\vglue 0.3cm
{\rightskip=3pc
 \leftskip=3pc
 \tenrm\baselineskip=12pt
 \noindent
This is a short review of present and future limits on 
new neutral gauge bosons, in particular on hadrophilic 
or leptophobic $Z'$s recently proposed to interpret the 
observed fluctuations of $\Gamma _{c,b}$ at LEP. Light 
gauge bosons coupled to lepton number differences or 
to baryon number are also examples of the model 
dependence of these bounds. The mixing between the 
$U(1)$ factors plays an important role in the 
phenomenology of these extended electroweak models. 
Future improvements based on the analysis of precise 
electroweak data are emphasized.}

\vglue 1.cm
{\centerline{\elevenbf 1. Introduction}}
\vglue 0.5cm

\elevenrm
There is a large literature on new weak interactions and 
their limits. For recent reviews on the 
discovery and identification of extra gauge bosons see 
Ref. \cite{CGH,LEP}. (See Refs. \cite{OLDREVIEWS,Fa} 
for earlier reports.)

Up to now there is no compelling evidence for a new $Z'$. 
On the contrary the standard model (SM) is in agreement 
with present data \cite{SM,Al,La}, although the observed 
fluctuations in the charm and bottom $Z$ widths, 
$\Gamma _{c,b}$, at LEP have led to speculate on the 
possibility of a new hadrophilic or leptophobic gauge 
boson \cite{LEP,CAB}. At any rate many extensions of the 
SM predict new gauge bosons. As on the other hand present 
limits on new interactions are rather weak and very model 
dependent, it is interesting to study the possibility of 
their discovery at future colliders and to compare the 
$Z'$ physics potential of the different machines. 

We present a short review of five topics. 
Present \cite{La,CAB} and future \cite{CGH,LEP} $Z'$ 
limits are discussed in Section 2, including the 
hadrophilic or leptophobic models proposed to explain 
the $\Gamma _{c,b}$ deviations. In order to further 
show the model dependence of these bounds, in Section 3 
we give two examples of extra interactions with gauge 
boson masses $M_{Z'}$ near or below the $Z$ mass. 
These models, which are interesting by themselves, 
gauge lepton number differences \cite{emt} 
and baryon number \cite{CM}, respectively. 
The latter is also hadrophilic or leptophobic but it has 
not the correct charge assignments to accommodate the 
observed fluctuations of $\Gamma _{c,b}$.
The mixing between the abelian subgroups $U(1)_Y$ 
and $U(1)_B$, which plays a major role in the 
phenomenology of the model, is studied in some detail 
\cite{HA}. 
Finally, in Section 4 we revise the effects of 
radiative corrections in models with an extra $U(1)$ 
factor \cite{AMP}. 

\vglue 1.cm
{\centerline{\elevenbf 2. $Z'$ limits}}
\vglue 0.5cm

An extra abelian gauge interaction is parametrized by the 
mass of the new gauge boson $Z'$, $M_{Z'}$, its mixing 
with the standard $Z^0$ boson, $\theta _3$ ($= - \phi, 
- \theta$ in Refs. \cite{CGH,La}, respectively), and 
its couplings with the known fermions. (If the model 
requires extra matter there are more ($Z'$) couplings 
whose effects on the processes involving the observed 
fermions are mainly encoded in the total $Z'$ width, 
$\Gamma _{Z'}$.) Many of these parameters must be fixed 
when comparing with experiment in order to obtain 
sensible results. The ${Z^{0}}'$ couplings are 
usually fixed, leaving the $Z'$ mass (and the $Z'Z^0$ 
mixing angle) free. Thus, limits are only quoted for 
definite models, being very model dependent. 

\rightskip=0pc
\leftskip=0pc
\vskip 0.5cm
{\centerline{\elevenit 2.1 Present $Z'$ limits}}
\vskip 0.5cm

Present bounds come from lepton pair production at 
TEVATRON (direct) and from precise electroweak data 
(indirect). These constraints are usually illustrated 
giving the limits for five specific models, $\chi , 
\psi , \eta , LR, SSM$. The $\chi , \psi , \eta , LR$ 
models can be embedded in $E_6$, whereas the $SSM$ model 
assumes a new gauge boson with the same couplings as 
but heavier than the standard $Z^0$. We collect in 
Table 1 the bounds on $M_{Z'}$ for these models 
(see Refs. \cite{CGH,La}). The integrated luminosity at 
TEVATRON is $19.6\ pb^{-1}$ (lepton pairs include 
$e^+e^-, \mu ^+\mu ^-$) and the global electroweak 
analysis (requiring $95\ \% \ C.L.$) includes the 1993 
LEP data. The constrained limits correspond to minimal 
Higgs contents. The direct bounds will improve with the 
integrated luminosity at TEVATRON. If no signal is 
observed for $70\ pb^{-1}$, the limits will increase 
$\sim 120\ GeV$ \cite{CGH,AMQ}. 
All these limits are obtained assuming 
that the new gauge boson can only decay into known 
fermions. If the open channels include the three complete 
$E_6$ supersymmetric families, the $M_{Z'}$ limits are 
reduced by $50-100\ GeV$, depending on the model 
\cite{CGH,AMQ}. The indirect bounds will improve slightly 
when all LEP data are analysed. This is assuming that no 
significative departure from the SM is found. 

\vglue 0.2cm
\begin{center}
\begin{tabular}{|c|ccc|}
\hline
\multicolumn{1}{|c}{} & direct & $\begin{array}{c} 
{\rm indirect} \\ {\rm (unconstrained)} \\ \end{array}$ & 
\multicolumn{1}{c|}{$\begin{array}{c} {\rm indirect} \\ 
{\rm (constrained)} \\ \end{array}$} \\
\hline
$\chi $ & 425 & 330 & 920 \\ 
$\psi $ & 415 & 170 & 170 \\  
$\eta $ & 440 & 220 & 610 \\  
$LR$ & 445 & 390 & 1360 \\  
$SSM$ & 505 & 960 &  \\  
\hline
\end{tabular}
\end{center}
\vglue 0.1cm

\vglue 0.1cm
\rightskip=3pc
\leftskip=3pc
{\tenrm \baselineskip=12pt 
\noindent
Table 1. Present $M_{Z'}$ limits (in GeV) for typical models.}
\vglue 0.1cm 

\rightskip=0pc
\leftskip=0pc
\vglue 0.2cm

At present there are fluctuations in $\Gamma _{c,b}$ 
at LEP. The models above do not explain these fluctuations. 
However with a different choice of $Z'$ charges this 
departure from the SM can be related to the existence 
of a new gauge boson. Several groups have discussed 
this possibility recently \cite{CAB}. The new $Z'$ must 
couple only to quarks (hadrophilic) 
for the $Z$ couplings to leptons are in very good 
agreement with the SM values. In the first two papers 
of Ref. \cite{CAB} the 
new quark charge ratios are fixed to accommodate the 
$\Gamma _{c,b}$ fluctuations; whereas in the third one 
it is pointed out that the unique $U(1)$ in $E_6$ 
with zero lepton charges (leptophobic), 
$Y_{\eta}+\frac{1}{3}Y$ \cite{AQZ} ($Y_{\eta}=-Y'$ in 
this reference), 
can be effective at the electroweak scale 
if the necessary $U(1)_Y\times U(1)_{\eta}$ mixing is 
generated when renormalizing the gauge couplings down 
to low energies. As a matter of fact this relatively 
large mixing can be  
obtained for a definite matter content. The quark 
couplings also explain the observed 
$\Gamma _{c,b}$ deviations. 

\rightskip=0pc
\leftskip=0pc
\vskip 0.5cm
{\centerline{\elevenit 2.2 Future $Z'$ limits}}
\vskip 0.5cm

At future colliders the $Z'$ limits will improve if no 
departure from the SM predictions is observed. In 
Table 2 we gather the expected bounds at present and 
future colliders. A large part of the detailed discussion 
of how to derive these bounds is presented in Ref. 
\cite{CGH} (see also Ref. \cite{ACL}), and we refer 
the interested reader to this review. 

\vglue 0.2cm
\begin{center}
\begin{tabular}{|c|ccccc|}
\hline
\multicolumn{1}{|c|}{Collider} & TEVATRON & LHC & 
LEP200 & NLC & 
\multicolumn{1}{c|}{HERA} \\
\hline
$\chi $ & 775 & 3040 & 695 & 3340 & 235 \\ 
$\psi $ & 775 & 2910 & 269 & 978 & 125 \\  
$\eta $ & 795 & 2980 & 431 & 1990 & 215 \\  
$LR$ & 825 & 3150 & 493 & 2560 & 495 \\  
\hline
\end{tabular}
\end{center}
\vglue 0.1cm

\vglue 0.1cm
\rightskip=3pc
\leftskip=3pc
{\tenrm \baselineskip=12pt 
\noindent
Table 2. $M_{Z'}$ limits (in GeV) for typical models 
at future colliders. TEVATRON ($p\bar p$ at $\sqrt s = 1.8 
\ TeV, \ {\cal L} _{int} = 1\ fb^{-1}$), LEP200 
($e^+e^-$ at $\sqrt s = 0.2\ TeV, \ {\cal L} _{int} = 0.5
\ fb^{-1}$) and HERA ($ep$ at $\sqrt s = 0.314\ TeV, \ 
{\cal L} _{int} = 0.6\ fb^{-1}$) provide a modest 
improvement compared to LHC ($pp$ at $\sqrt s = 10\ TeV, 
\ {\cal L} _{int} = 40\ fb^{-1}$) and NLC ($e^+e^-$ at 
$\sqrt s = 0.5\ TeV, \ {\cal L} _{int} = 50\ fb^{-1}$).}
\vglue 0.1cm 

\rightskip=0pc
\leftskip=0pc
\vglue 0.2cm

\vglue 1.cm
{\centerline{\elevenbf 3. Leptonic and baryonic 
extended gauge models}}
\vglue 0.5cm

In this Section we discuss two classes of models which 
evade the usual $M_{Z'}$ limits. In both cases $M_{Z'}$ 
can be smaller than $M_Z$. One simplifying but strong 
assumption in many analyses is that the extra 
interaction is universal. Below we comment on models 
gauging lepton number differences as examples of 
non-universal interactions. The class of models gauging 
baryon number is also reviewed. This interaction is 
universal but provides an example of models where the 
mixing between $U(1)$ factors is phenomenologically 
relevant. The leptophobic model $Y_{\eta}+\frac{1}{3}Y$ 
above is another example of hadrophilic model evading 
usual $M_{Z'}$ bounds. 

\rightskip=0pc
\leftskip=0pc
\vskip 0.5cm
{\centerline{\elevenit 3.1 Lepton number difference 
extensions}}
\vskip 0.5cm

These models result from requiring that the extra $U(1)$ 
be anomaly free without adding new fermions. The lepton 
number differences, $L_e-L_{\mu},\ L_e-L_{\tau},\ 
L_{\mu}-L_{\tau}$, are the only solutions \cite{emt}, 
up to fermion mixing \cite{DJV}. Quark charges are zero. 
The constraints on these models are not very stringent, 
obviously it is difficult to constrain 
$L_{\mu}-L_{\tau}$. $e^+e^-$ colliders constrain the other 
two lepton differences. For example LEP can set a bound 
on the $L_e-L_{\tau}$ gauge boson mass $\sim 130 
\frac{g'}{e}\ GeV$ \cite{ABR}.

\rightskip=0pc
\leftskip=0pc
\vskip 0.5cm
{\centerline{\elevenit 3.2 Baryon number gauge 
models}}
\vskip 0.5cm

The experimental limits on extra interactions involving  
only quarks are somewhat weak. A class of models 
with a new $Z'$ coupling only to baryon number is 
phenomenologically allowed, even if $M_{Z'} < M_Z$ and 
the new gauge coupling strength is order 1 \cite{CM}. This 
extra $Z'$ couples universally to the three known families 
but the matter content of the model is enlarged with heavy 
fermions to render the model anomaly-free. 
Although the $Z'$ couplings to leptons are zero at tree 
level, small couplings are generated at higher orders. 
As a matter of fact it is the non-vanishing of these couplings 
what may allow for observing this new gauge boson in the 
dilepton channel at large hadron colliders. The origin of 
these small couplings is the mixing of the abelian kinetic 
terms. 

Let us discuss this mixing more carefully for it is one 
example of the general case examined below. In the absence 
of a symmetry, all $U(1)$ factors of a gauge group mix at 
some order in perturbation theory \cite{AMP}. This mixing 
may be required in order to renormalize the 
theory or can be generated at higher orders as in this class 
of models. The new contribution can be written as an off-diagonal 
kinetic term which mixes the $U(1)$ gauge fields or in a canonical 
way, redefining the gauge fields in order to keep their 
kinetic terms diagonal and conventionally normalized. In 
the latter case the mixing appears in the $Z'$ current, 
which is modified by the addition of a mixing term 
proportional to the standard abelian current.

Thus, if $SU(3)_C\times SU(2)_L
\times U(1)_Y\times U(1)_B$ is unbroken, the new effects 
can be taken into account by adding an off-diagonal kinetic 
term to the canonical lagrangian for the $U(1)$ fields

\begin{equation}
{\cal L}_{kin}=-{\frac{1}{4}} \left(
F^Y_{\mu \nu}F^{Y \mu \nu}
+ 2cF^Y_{\mu \nu}F^{B \mu \nu} + 
F^B_{\mu \nu}F^{B \mu \nu}
\right), 
\end{equation}

\noindent
with $c$ a calculable quantity obtained by evolving the 
gauge couplings down to the electroweak scale. This 
lagrangian is equivalent to a canonical one with gauge 
fields 

\begin{equation}
\left( \begin{array}{c} A'^Y \\ A'^B \\ \end{array} 
\right) = 
\left( \begin{array}{cc}
1 & c \\ 0 & \sqrt {1-c^2} \\ \end{array}
\right) \left( \begin{array}{c}
A^Y \\ A^B \\ \end{array} \right)
\end{equation}

\noindent
and covariant derivatives 

\begin{equation}
D_{\mu} f_i = \partial _{\mu} f_i + i 
(\tilde q'^Y_i \ \tilde q'^B_i) \left( \begin{array}{c}
A'^Y_{\mu} \\ A'^B_{\mu} \\ \end{array} \right) f_i, 
\end{equation}

\noindent
where $f_i$ is a matter field with gauge couplings 

\begin{equation}
(\tilde q'^Y_i \ \tilde q'^B_i) = 
(q_i^Y \ q_i^B)
\left( \begin{array}{cc}
g^Y & 0 \\ 0 & g^B \\ \end{array} \right) 
\left( \begin{array}{cc}
1 & -\frac {c}{\sqrt {1-c^2}} \\ 0 & 
\frac {1}{\sqrt {1-c^2}} \\ \end{array}
\right) =
(q_i^Y \ q_i^B)
\left( \begin{array}{cc}
g^Y & -\frac {g^Yc}{\sqrt {1-c^2}} \\ 0 & 
\frac {g^B}{\sqrt {1-c^2}} \\ \end{array}
\right).
\end{equation}

\noindent 
Hence, $A'^Y$ will be identified with the hypercharge and 
$A'^B$ with a new gauge boson with charges 
$
\tilde q'^B_i = 
\frac{g^B}{\sqrt {1-c^2}} q_i^B - 
\frac{g^Y c}{\sqrt {1-c^2}} q_i^Y.  
$
This modified charges include, as the only new effect, 
a mixing term proportional to the 
hypercharge. It is apparent from the second formulation 
that new physical effects will require the 
exchange of the new gauge boson (for in the unbroken case 
there is no mass mixing) 
and that they are at least 
suppressed by a factor $c$ if they 
involve leptons, for which $q_i^B = 0$. 

After spontaneous symmetry breaking the rotation which 
diagonalizes the vector boson mass matrix introduces 
off-diagonal kinetic terms for the photon 
$A^{\gamma}_{\mu} = c_WA^Y_{\mu}+s_WA^W_{\mu}$ 
and the standard model gauge boson $A^{Z^0}_{\mu}= 
-s_WA^Y_{\mu}+c_WA^W_{\mu}$ in the non-canonical 
lagrangian (see Eq. (1)), 

\begin{equation}
{\cal L}_{kin} = -{\frac{1}{4}} 
( F^{\gamma}_{\mu \nu}F^{\gamma \mu \nu}
+ F^{Z^0}_{\mu \nu}F^{Z^0 \mu \nu} 
+ 2cc_W F^{\gamma}_{\mu \nu}F^{B \mu \nu}
- 2cs_W F^{Z^0}_{\mu \nu}F^{B \mu \nu} + 
F^B_{\mu \nu}F^{B \mu \nu} ), 
\end{equation}

\noindent
with $s_W (c_W) = sin \theta _W (cos \theta _W)$ the 
electroweak mixing. In this basis the boson mass matrix 
is diagonal because the vev $v$ 
giving a mass to $A^B$ transforms trivially under 
$SU(2)_L\times U(1)_Y$ and the standard Higgs doublet 
has zero $U(1)_B$ charge. However the eigenvalues are 
not the physical masses for the kinetic lagrangian is 
non-canonical. In the alternative formulation the 
lagrangian with canonical kinetic 
terms and a diagonal mass matrix has gauge couplings  

\begin{equation}
\begin{array}{c}
(\tilde q^{\gamma}_i \ \tilde q^Z_i \ \tilde q^{Z'}_i) = 
(q_i^W q_i^Y \ q_i^B)
\left( \begin{array}{ccc}
g^W & 0 & 0 \\ 0 & 
g^Y & -\frac {g^Yc}{\sqrt {1-c^2}} \\ 
0 & 0 & \frac {g^B}{\sqrt {1-c^2}} \\
\end{array} \right) 
\left( \begin{array}{ccc}
s_W & c_W & 0 \\ c_W & 
-s_W & 0 \\ 
0 & 0 & 1 \\
\end{array} \right)
\left( \begin{array}{ccc}
1 & 0 & 0 \\ 0 & 
c_3 & s_3 \\ 
0 & -s_3 & c_3 \\
\end{array} \right) \\ 
= (q_i^W q_i^Y \ q_i^B) 
\left( \begin{array}{ccc}
e & \frac {ec_W}{s_W}c_3 & \frac {ec_W}{s_W}s_3 \\ 
\sqrt {\frac {5}{3}}e & \sqrt {\frac {5}{3}}
\frac {e}{c_W}(-s_Wc_3+\frac{cs_3}{\sqrt {1-c^2}})
& \sqrt {\frac {5}{3}}
\frac {e}{c_W}(-s_Ws_3-\frac{cc_3}{\sqrt {1-c^2}}) \\ 
0 & -\frac {g^Bs_3}{\sqrt {1-c^2}} 
& \frac {g^Bc_3}{\sqrt {1-c^2}} \\ 
\end{array} \right), 
\end{array}
\end{equation}

\noindent
with $e=g^Ws_W$ and $s_W=\frac{\sqrt 3 g^Y}
{\sqrt{3g^{Y 2}+5g^{W 2}}}$. The $Z'Z^0$ mixing 
$s_3(c_3)=sin\theta _3(cos\theta _3)$  
is a function of the gauge boson masses, $s_3=
sign(c)\sqrt {\frac{M_{Z^0}^2-M_Z^2}
{M_{Z'}^2-M_Z^2}}$ \cite{AQZ}. 

The relevance of $c_{\gamma}=cc_W$ and $c_Z=-cs_W$ 
and the phenomenological signatures of the model are 
worked out in detail in Ref. \cite{CM}. In the canonical 
basis, with canonical kinetic lagrangian and physical 
vector boson masses, the standard model with an extra 
$U(1)$ is in general described by a $3\times 3$ gauge 
coupling matrix, product of a triangular matrix with zero 
$WY$ and $WB$ entries (as required by $SU(2)_L$ 
invariance) by the rotation matrix diagonalizing the 
gauge boson mass matrix, which only depends on two 
mixing angles, $\theta _W$ and $\theta _3$. 
A non-zero $\theta _3$ modifies the $\rho$ parameter 
and the $Z$ couplings to fermions \cite{AQZ}. 

\vglue 1.cm
{\centerline{\elevenbf 4. Radiative corrections 
in extended electroweak models}}
\vglue 0.5cm

Precise limits on the SM parameters are obtained 
comparing with one-loop expressions \cite{CERNY}.
The corresponding equations for 
$SU(3)_C\times SU(2)_L\times U(1)_Y\times U(1)'$
have not been worked out. New one-loop 
corrections can be sometimes neglected.  
This approximation improves when the effective 
strength of the new force $\sim \frac{g'}{M_{Z'}}$ 
decreases. At any rate experiment allows only for 
small departures of the SM \cite{SM}. 
However, to obtain limits on new interactions new 
one-loop corrections would have to be included if the 
relevant range of $Z'$ masses (couplings) is low (large) 
enough. 

It is important in any case to use 
a general parametrization of the new interaction. 
This guarantees a consistent expansion in perturbation 
theory and justifies the use of approximate tree-level 
expressions in the fits \cite{AMP,DS}. 
All new free parameters must be included in the analysis.  
Even if some are small (and generated by 
renormalization of the high energy couplings 
down to the electroweak scale). 
Two recent examples, where this is crucial, have been 
already discussed in the previous Sections. Both are 
related to the possible mixing between 
$U(1)$ factors. In the 
leptophobic model in Section 2 \cite{CAB} this mixing 
is essential to vanish the $Z'$ couplings to leptons 
at the electroweak scale; whereas in the model 
gauging baryon number this mixing is responsible 
of the eventual observation of the new gauge boson 
in lepton pair production \cite{CM}.
\vglue 0.8cm

{\elevenbf\noindent Acknowledgements \hfil}
\vglue 0.3cm

This work was partially supported by CICYT under contract 
AEN94-0936, by the Junta de Andaluc\'\i a and by the 
European Union under contract CHRX-CT92-0004. 
We thank F. Cornet and C. Verzegnassi for useful comments 
and the organizers of the School for their warm 
hospitality. 

\vglue 1.cm

{\elevenbf\noindent References \hfil}
\vglue 0.3cm


\begin{thebibliography}{9}

\bibitem{CGH} M. Cveti\v c and S. Godfrey, UPR-648-T, hep-ph/9504216, 
summary of the Working Subgroup on Extra Gauge Bosons of the DPF 
long-range planning study to be published in {\elevenit Electro-weak Symmetry 
Breaking and Beyond the Standard Model}, Eds. T. Barklow, S. Dawson, H. Haber 
and J. Seigrist (World Scientific 1995);
J.L. Hewett, SLAC-PUB-95-6960, hep-ph/9507400, in {\elevenit 10th Topical 
Workshop on Proton-Antiproton Collider Physics}, Batavia, IL, 9-13 May, 
1995. 

\bibitem{LEP} P. Chiappetta et al., 
Proceed. of LEP2 workshop (1995), CERN report in preparation.

\bibitem{OLDREVIEWS} F. Zwirner, {\elevenit Int. J. Mod. Phys.}
{\elevenbf A3} (1988) 49;
J.L. Hewett and T.G. Rizzo, {\elevenit Phys. Rep.} {\elevenbf 183}
(1989) 193;
see also P. Langacker and M. Luo, 
{\elevenit Phys. Rev.} {\elevenbf D44} (1991) 817, and 
references there in.

\bibitem{Fa} F. del Aguila, {\elevenit Acta Physica Polonica}
{\elevenbf B25} (1994) 1317.

\bibitem{SM} D. Schaile, {\elevenit XXVII Int. Con. on HEP}, 
Glashow, UK, 20-27 July 1994, Eds. P.J. Bussey and 
I.G. Knowles, p. 27; 
M. Mart\'\i nez, IFAE-UAB/95-01, Lectures given at 
{\elevenit Frontieres in Particle Physics}, Institut 
d'Etudes Scientifiques de Cargese, Corsica (France), 
August 1994; 
The LEP Collaborations ALEPH, DELPHI, L3, OPAL and the 
LEP Electroweak Working Group, CERN-PPE preprint, in 
preparation. 

\bibitem{Al} G. Altarelli, CERN-TH. 7464/94, Summary Talk 
given at the {\elevenit Tennessee International Symposium on 
Radiative Corrections}, Gatlinburg, USA, July 1994.

\bibitem{La} P. Langacker, hep-ph/9412361, to be published 
in {\elevenit Precision Tests of the Standard Electroweak Model}, 
Ed. P. Langacker (World Scientific, Singapore 1994).

\bibitem{CAB} P. Chiappetta et al., CPT-96/P.3304, 
hep-ph/9601306; G. Altarelli et al., CERN-TH/96-20, 
hep-ph/9601324; K.S. Babu, C. Kolda and J. March-Russell, 
IASSNS-HEP-96/20, hep-ph/9603212; for earlier work 
B. Holdom, {\elevenit Phys. Lett.} {\elevenbf B339} 
(1994) 114; {\elevenbf B351} (1995) 279.

\bibitem{emt} X.G. He, G.C. Joshi, H. Lew and R.R. Volkas, 
{\elevenit Phys. Rev.} {\elevenbf D44} (1991) 2118. 

\bibitem{CM} C.D. Carone and H. Murayama, {\elevenit Phys. 
Rev. Lett.} {\elevenbf 74} (1995) 3122; {\elevenit Phys. 
Rev.} {\elevenbf D52} (1995) 484; A. Nelson and 
N. Tetradis, {\elevenit Phys. Lett.} {\elevenbf B221} 
(1989) 80. 

\bibitem{HA} B. Holdom, {\elevenit Phys. Lett.} 
{\elevenbf B166} (1986) 196; 
F. del Aguila et al., 
{\elevenit Nucl. Phys.} {\elevenbf B283} (1987) 50; 
F. del Aguila, G. Coughlan and M. Quir\'os, 
{\elevenit Nucl. Phys.} {\elevenbf B307} (1988) 633;
{\elevenit Erratum} {\elevenbf B312} (1989) 751; see 
also B. Holdom, {\elevenit Phys. Lett.} 
{\elevenbf B259} (1991) 329.

\bibitem{AMP} F. del Aguila, M. Masip and M. P\'erez-Victoria, 
{\elevenit Nucl. Phys.} {\elevenbf B456} (1995) 531. 

\bibitem{AMQ} F. del Aguila, J.M. Moreno and M. Quir\'os, 
{\elevenit Phys. Rev.} {\elevenbf D40} (1989) 2481; 
{\elevenbf D41} (1990) 134; 
{\elevenit Erratum} {\elevenbf D42} (1990) 262. 

\bibitem{AQZ} F. del Aguila, M. Quir\'os and F. Zwirner, 
{\elevenit Nucl. Phys.} {\elevenbf B287} (1987) 419. 

\bibitem{ACL} M. Cveti\v c and P. Langacker, 
{\elevenit Phys. Rev.} {\elevenbf D46} (1992) 4943; 
T. Rizzo, {\elevenit Phys. Rev.} {\elevenbf D47} 
(1993) 965; 
F. del Aguila, M. Cveti\v c, and P. Langacker, 
{\elevenit Phys. Rev.} {\elevenbf D48} (1993) R969; 
A. Djouadi, A. Leike, T. Riemann, D. Schaile and C. 
Verzegnassi, 
{\elevenit Z. Phys.} {\elevenbf C56} (1992) 289; 
F. del Aguila and M. Cveti\v c, 
{\elevenit Phys. Rev.} {\elevenbf D50} (1994) 3158;
A. Leike, {\elevenit Z. Phys.} {\elevenbf C62} (1994) 265; 
D. Choudhury, F. Cuypers and A. Leike, 
{\elevenit Phys. Lett.} {\elevenbf B333} (1994) 531; 
F. del Aguila, M. Cveti\v c, and P. Langacker, 
{\elevenit Phys. Rev.} {\elevenbf D52} (1995) 37. 

\bibitem{DJV} G. Dutta, A.S. Joshipura and K.B. Vijaykumar, 
{\elevenit Phys. Rev.} {\elevenbf D50} (1994) 2109. 

\bibitem{ABR} F. del Aguila, J. Bernab\'eu and N. Rius, 
{\elevenit Phys. Lett.} {\elevenbf B280} (1992) 319. 

\bibitem{CERNY} Reports of the Working Group on Precision 
Calculations for the Z Resonance, CERN 95-03, Eds. D. Bardin, 
W. Hollik and G. Passarino. 

\bibitem{DS} G. Degrassi and A. Sirlin, 
{\elevenit Phys. Rev.} {\elevenbf D40} 
(1989) 3066; 
L. Lavoura, {\elevenit Phys. Rev.} {\elevenbf D48} 
(1993) 2356. 


\end{thebibliography}
\end{document}